\documentclass{sig-alternate-05-2015}

\usepackage{listings}
\usepackage{hyperref}
\usepackage{afterpage}
\usepackage{graphicx}
\usepackage{color}
\usepackage{xspace,framed}
\usepackage{tabularx}
\usepackage{subfigure}

\DeclareGraphicsExtensions{.pdf,.png,.jpg}

\begin{document}

\CopyrightYear{2017}
\setcopyright{acmcopyright}
\conferenceinfo{}{}
\isbn{}
\doi{}



\newcommand{\commentIB}[1]{\textcolor{blue}{[IB:#1]}}
\newcommand{\highlight}[1]{\textcolor{red}{#1}}

\newcommand\tool{{DSVerifier Toolbox}\xspace}
\newcommand{\fwl}[1]{\mathcal{FWL}[#1]}

\title{Verifying Digital Systems with MATLAB}

\numberofauthors{4}
%
\author{
%
%
\alignauthor
Lennon Chaves and \\ Iury Bessa\\
       \affaddr{Federal University of Amazonas}\\
       \texttt{lennonchaves@ufam.edu.br} \\
			\texttt{iurybessa@ufam.edu.br}
\alignauthor
Lucas Cordeiro and \\ Daniel Kroening \\
       \affaddr{University of Oxford}\\
       \texttt{lucas.cordeiro@cs.ox.ac.uk}
       \texttt{kroening@cs.ox.ac.uk}
\alignauthor
Eddie Filho\\
       \affaddr{Samsung Electronics}\\
       \texttt{eddie.l@samsung.com}              
}
\maketitle
\begin{abstract}
A MATLAB toolbox is presented, with the goal of checking occurrences of design errors typically found in fixed-point digital systems, considering finite word-length effects. In particular, the present toolbox works as a front-end to a recently introduced verification tool, known as Digital-System Verifier, and checks overflow, limit cycle, quantization, stability, and minimum phase errors, in digital systems represented by transfer-function and state-space equations. It provides a command-line version, with simplified access to specific functions, and a graphical-user interface, which was developed as a MATLAB application. The resulting toolbox is important for the verification community, since it shows the applicability of verification to real-world systems.
\end{abstract}

%
%


\printccsdesc


\section{Introduction}

Currently, digital systems ({\it e.g.}, filters and controllers) are used in a wide variety of applications, due to some advantages over their analog counterparts, such as reliability, flexibility, and cost. Nonetheless, there are disadvantages regarding their use: since they are normally implemented in microprocessors, errors might be introduced, due to quantization and related round-off effects~\cite{diniz}.

Hardware choice, structure representations ({\it e.g.}, direct forms), and implementation features ({\it e.g.}, number of integer and fractional bits, in fixed-point arithmetic) can heavily influence a given digital-system's precision and performance~\cite{daes20161}. Additionally, such implementations are particularly susceptible to finite word-length (FWL) effects (e.g., overflows, limit cycles, and poles and zeros sensitivity), which have the potential to reduce the associated reliability and efficiency. Previous studies have already shown that FWL effects might lead to excessive power loss and lifespan reduction, in power converters~\cite{peterchev} and oscillators~\cite{peretzLC_ressonant}; they might also affect the stability and performance of feedback control systems~\cite{bhattacharyya}. Thus, it is important to develop techniques that provide proof of correctness and safety, regarding digital-system implementations affected by FWL effects.

In order to detect the mentioned errors in digital systems, a model-checking procedure based on Boolean Satisfiability (SAT) and Satisfiability Modulo Theories (SMT) has been proposed, named as Digital-System Verifier (DSVerifier)~\cite{spindsverifier}. DSVerifier checks specific properties related to overflow, limit cycle, stability, and minimum-phase, in digital-system implementations~\cite{daes20161}, and also supports the verification of robust stability, considering parametric uncertainties for closed-loop systems represented by transfer functions~\cite{Bessa16}. Recently, DSVerifier was extended to support state-space systems, considering single-input single-output (SISO) and multiple-input multiple-output (MIMO) systems~\cite{monteiro2016}, in order to verify violations in stability, controllability, observability, and quantization-error properties. Although those contributions present important advances regarding formal verification of digital systems, they do not offer any compatibility with tools usually employed in the design of digital filters and controllers ({\it e.g.}, MATLAB~\cite{matlab-toolbox}).

Currently, there are several toolboxes in MATLAB with functions to facilitate the design of digital systems~\cite{matlab-toolbox}. For instance, the fixed-point designer toolbox provides data-types and tools for developing fixed-point digital systems. There are also other modules with different objectives, such as optimization, control systems, and digital signal processing. In particular, users could employ formal verification methods to identify errors and generate test vectors for reproducing errors. In that sense, Simulink Design Verifier~\cite{matlab-toolbox} employs formal methods to identify hidden design errors, without extensive simulation runs; it detects blocks that result in integer overflow, dead logic, array access, division by zero, and requirement violations. Additionally, it is possible to use tools for detecting and proving errors in source code written in C/C++, through Polyspace Bug Finder~\cite{matlab-toolbox}. Nonetheless, both tools are unable to automatically detect specific errors related to digital system design ({\it e.g.}, limit cycle, stability, and minimum-phase), unless an engineer provides additional assertions to be checked~\cite{dsvalidator}. Finally, the mentioned tools do not consider FWL effects during verification, and also, there is no MATLAB toolbox for verifying digital systems using symbolic model checking based on SAT and SMT solvers. 

The present paper addresses this problem and presents a MATLAB toolbox for DSVerifier,\footnote{Available at http://www.dsverifier.org} known as \tool, which applies SAT- and SMT-based model checking to digital systems~\cite{spindsverifier}, in the MATLAB's environment. The main advantage regarding the use of a MATLAB toolbox lies on designing digital systems in MATLAB and then promptly verifying their desired properties: overflow, limit-cycle, stability, minimum-phase, controlability, observability, and quantization error. Additionally, when using the DSVerifier Toolbox, an engineer is able to design a digital system with MATLAB, through tranfer-function or state-space representations and considering low-level systems parameters (implementation characteristics and numeric format), define realization forms ({\it e.g.}, delta and direct forms), and evaluate different overflow modes (wrap-around or saturate mode). Finally, if a verification procedure fails, the \tool returns a counterexample in a ``.MAT'' file, which explores the violation, considering inputs, initial states, and outputs, in order to reproduce a counterexample.

\section{DSVerifier-Aided Verification \\ Methodology}
\label{dsv-engine}

The proposed verification methodology is based on DSVerifier and can be split into four steps. In step $1$, a digital system is designed (in open- or closed-loop), with any design technique or tool. Later, implementation features are defined in step $2$, {\it i.e.}, FWL format (number of bits in the integer and fractional parts), dynamic range, and realization form (direct or delta). DSVerifier formulates a FWL function $\fwl{\cdot}:\mathcal{P}^{n}\rightarrow\mathcal{P}^{n}$, where $\mathcal{P}^{n}$ is a space of polynomials of $n$-th order, in order to reproduce the effects of the chosen FWL format over the coefficients of a digital system. $\fwl{A(z)}$ is the  polynomial $A(z)$ with FWL effects that is used to compute round-off effects, in digital systems. Those definitions are then passed to DSVerifier, along with hardware specifications and other verification parameters ({\it e.g.}, verification time) and properties to be checked. 

In particular, with respect to open-loop systems in transfer-function representation, DSVerifier supports verification of overflow, stability, minimum-phase, limit cycle, and quantization error properties, while it provides verification for stability, controllability, observability, and quantization properties, in state-space representation. Regarding closed-loop systems represented by a controller and a plant in transfer function form, DSVerifier is able to verify stability, quantization error, and limit-cycle, while it checks stability, controllability, observability, and quantization error, when state-space equations are employed. 

Once the configuration has been set up in step $3$, the verification process is then started in step $4$, with the chosen model-checking tool (CBMC~\cite{cbmc} or ESBMC~\cite{esbmc} can be used as back-end). DSVerifier then checks the desired properties and returns ``successful'', if there is no property violation in the proposed implementation, or ``failed'' together with a counterexample, which contains inputs and states that lead the system under evaluation to a given property violation. The implementation features and design should then be improved, based on the available counterexample, {\it i.e.}, realization, representation, and FWL format can then be re-chosen, in order to avoid errors. Finally, such a process is repeated until a digital controller implementation does not present any failure.

\tool uses bounded model checking (BMC) as verification engine. The basic idea of the BMC technique is to check the negation of a given property, at a given depth. Given a transition system $M$, a property $\phi$, and a bound $k$, the employed verification engine unrolls the transition system $k$ times and translates it into a verification condition $\psi$, in such a way that $\psi$ is satisfiable if and only if $\phi$ has a counterexample of depth less than or equal to $k$. Thus, overflow, limit cycle, and quantization errors, in transfer-function representation, and quantization error verification, when employing a state-space representation must be unrolled $k$ times, in order to find violations (verification is incomplete, but sound up to $k$). In contrast, properties such as stability and minimum-phase, in transfer-function representation, and controllability, stability, and observability, in state-space representation, do not need a definition of $k$ (verification is complete and sound).

\section{Verifying Digital Systems with DSVerifier Toolbox}


\subsection{The Employed Verification Methodology}

\begin{figure}[ht!]
\centering
\includegraphics[width=0.6\textwidth]{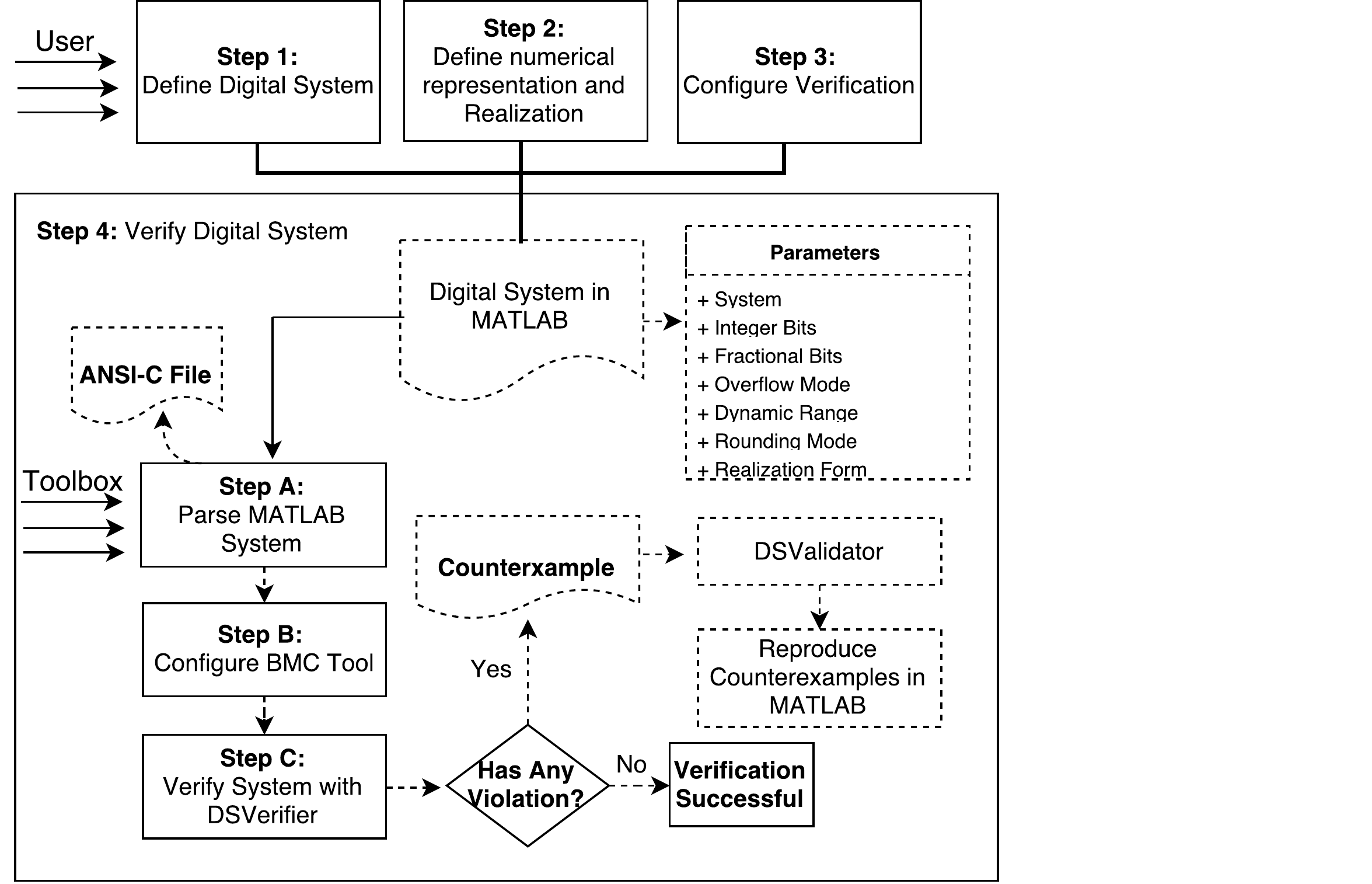}
\caption{DSVerifier Toolbox's Verification Methodology.}
\label{method_toolbox}
\end{figure}

Fig.~\ref{method_toolbox} shows the proposed DSVerifier toolbox's verification methodology, which can be split into two main stages: manual (user) and automated (toolbox) procedures. In the former, the user manually performs steps $1$ to $3$, which are the same tasks performed by DSVerifier (design of a digital-system, definition of numerical representation, realization form, and verification configuration). Note that all those specifications are provided as parameters (and translated to a struct format in the automated procedures performed by the toolbox), as can be seen in Fig.~\ref{method_toolbox}.

The toolbox's automated engine (steps $A$ to $C$) receives a digital system's specification (as parameters) and verifies the desired property $\phi$. In step $A$, intermediate ANSI-C code for the desired implementation is created, based on parameters that are then translated into a struct format (in MATLAB) and parsed, while the respective BMC tool is set and all requirements are configured in step $B$. Finally, in step $C$, the resulting ANSI-C code is passed to a highly-efficient BMC tool ({\it e.g.}, CBMC or ESBMC) and then converted into SAT or SMT formulae, which are checked by the respective solver. If any violation is found, then DSVerifier reports a counterexample, which contains system inputs that lead to a failure; otherwise, it returns a successful verification.
In particular, in case of a failure, the proposed toolbox receives a counterexample and generates a corresponding ``.MAT' file. It is worth noticing that the same counterexample could be reproduced and validated with the DSValidator~\cite{dsvalidator} tool.

\subsection{DSVerifier Toolbox Features}


\begin{enumerate}
    \item{\textbf{Digital-system representation:} \tool handles digital systems represented by transfer-function and state-space representations.}
    \item{\textbf{Realization:} \tool performs the verification of direct forms, such as direct-form I (DFI), direct-form II (DFII), and transposed direct-form II (TDFII), and also delta forms, such as delta direct-form I (DDFI), delta direct-form II (DDFII), and delta transposed direct-form II (TDDFII).}
    \item{\textbf{Properties:} \tool verifies, for transfer-function representation, stability, overflow, minimum phase, limit-cycle, and quantization error, while in state-space representation, it verifies stability, quantization error, observability, and controllability properties.}
     \item{\textbf{Closed-loop systems:} \tool verifies stability, limit-cycle, and quantization error in transfer-function representation, while for state-space systems, all properties mentioned for open-loop systems are checked, via state feedback matrix.}
     \item{\textbf{BMC tools:} \tool handles the verification for digital-systems using CBMC~\cite{cbmc} or ESBMC~\cite{esbmc} as back-end, in order to perform BMC.}
\end{enumerate}

\subsection{DSVerifier Toolbox Usage}

In order to explain the \tool's workflow, the following second-order controller for a A/C motor plant is used, which can be found in a set of benchmarks ({\it e.g.}, unmanned aerial vehicle) available online:\footnote{http://www.dsverifier.org/benchmarks}

\begin{equation}
\label{equation_controller}
H(z)=\frac{z^{3}-2.819z^{2}+2.6370z-0.8187}{z^{3}-1.97z^{2}+1.033z-0.06068}.
\end{equation}

\subsubsection{Command Line Version}

Currently, users must provide a digital system described as a MATLAB system, {\it i.e.}, using a \texttt{tf} (for transfer-function) or an \texttt{ss} (for state-space) command, in order to design systems. In this command-line version, \tool is invoked to check the desired property $\phi$ and digital system representation ({\it i.e.}, transfer-function, closed-loop, or state-space). Table~\ref{table-functions} shows the \tool's commands that perform the proposed verification and the required parameters for each property. In particular, a $k$ bound is required only in some properties, as mentioned in section~\ref{dsv-engine}. In Table~\ref{table-functions}, $system$ represents the digital system in transfer-function or state-space format, $intbits$ is the integer part, $fracbits$ is the fractional part, $max$ and $min$ are the maximum and minimum dynamic range, respectivelly, $bound$ is the $k$ bound to be employed during verification, $cmode$ is the connection mode, for closed-loop systems in transfer-function (series or feedback), and $error$ is the maximum possible value in the quantization error check.


%
\begin{table*}[t]
\centering
\scriptsize
\setlength\tabcolsep{3pt}
\caption{\tool's commands and parameters used during verification procedures.}
\label{table-functions}
\vline
\begin{tabular}{l|c|c|c|c|c|c|c|c|c|}
\hline
\textbf{Verification Command} & system & intbits & fracbits & max & min & bound & cmode & error \\
\hline                               
verifyStability & x & x & x & x & x & & & \\ \hline
verifyOverflow  & x & x & x & x & x & x & & \\ \hline
verifyError & x & x & x & x & x & x & & x\\ \hline
verifyMinimumPhase & x & x & x & x & x & & & \\ \hline
verifyLimitCycle & x & x & x & x & x & x & & \\ \hline
verifyClosedStability & x & x & x & x & x & & x & \\ \hline
verifyClosedQuantizationError & x & x & x & x & x & x & x & x \\ \hline
verifyClosedLimitCycle & x & x & x & x & x & x & x & \\ \hline
verifyStateSpaceStabiltiy & x & x & x &  &  &  &  &  \\ \hline
verifyStateSpaceControllability & x & x & x &  &  &  &  &  \\ \hline
verifyStateSpaceObservability & x & x & x &  &  &  &  & \\ \hline
verifyStateSpaceQuantizationError & x & x & x &  &  & x  &  & x \\
\hline
\end{tabular}
\end{table*}
Additionally, optional parameters can be included, such as overflow mode, rounding mode, BMC tool, solver, quantization error mode, delta coefficient (for delta realization), and other attributes that DSVerifier supports.\footnote{All functions implemented in \tool are detailed in the \href{http://dsverifier.org/dsverifier-toolbox}{Toolbox's Documentation}.} All available functions w.r.t. the \tool have been exhaustively tested and experimental results are available online.\footnote{http://www.dsverifier.org/benchmarks}

\subsubsection{Illustrative Example}

In order to illustrate the \tool's usage, Fig.~\ref{toolbox-usage} shows the stability verification for the digital system specified in Eq.~\ref{equation_controller}, using a fixed-point format $\left\langle 2,13\right\rangle$ and a dynamic range $[1,-1]$.

\begin{figure}[ht]
\scriptsize
\begin{lstlisting}[xleftmargin=.025\textwidth,frame=single,]
>> num = [1.0000 -2.8190 2.6370 -0.8187];
>> den = [1.0000 -1.9700 1.0330 -0.0607];
>> system = tf(num,den,0.001);
>>
>> verifyStability(system,2,13,1,-1);
>> VERIFICATION SUCCESSFUL
\end{lstlisting}
\vspace{-0.2cm}
\caption{Verifying Stability for a digital-system designed in MATLAB, with a fixed-point format  $\left\langle 2,13\right\rangle$.}
\label{toolbox-usage}
\end{figure}

If the fixed-point format is changed to $\left\langle 12,3\right\rangle$, for the same system described in Eq.~\ref{equation_controller}, the verification indicates a failure, {\it i.e.}, the digital system is unstable, as can be seen in Fig.~\ref{toolbox-unstable}, which indicates that the \tool is able to correctly verify digital systems with different implementations.
\begin{figure}[ht]
\scriptsize
\begin{lstlisting}[xleftmargin=.025\textwidth,frame=single,]
>> verifyStability(system,12,3,1,-1);
>> VERIFICATION FAILED
\end{lstlisting}
\vspace{-0.2cm}
\caption{Verifying Stability for a digital-system designed in MATLAB, with a fixed-point format  $\left\langle 12,3\right\rangle$.}
\label{toolbox-unstable}
\end{figure}

After verifying that the adopted digital system is unstable ({\it i.e.}, verification fails) with the fixed-point format $\left\langle 12,3\right\rangle$, the respective verification result can be confirmed by reproducing the counterexample generated by the \tool. As mentioned during the explanation of the proposed methodology, we can have a polynomial $A(z)$ with FWL effects, through the application of $\fwl{A(z)}$. In particular, we compute the roots of $\fwl{A(z)}$, in order to check stability. If any root has modulus equal or greater than one, then the system is unstable; otherwise, it is stable. When applying $\fwl{H(z)}$, with the first case ({\it i.e.}, $\left\langle 2,13\right\rangle$), and computing the roots of the denominator of $\fwl{H(z)}$, where $H(z)$ is introduced in Eq.~\ref{equation_controller} to represent a digital-system, we obtain the following set of poles: $I=\{0.9629, 0.9400, 0.0672\}$, from which one can conclude that all poles are inside the unit circle. This means that the mentioned system is stable, when the numeric representation $\left\langle 2,13\right\rangle$ is used; however, when applying $\fwl{H(z)}$ to the second case ({\it i.e.}, $\left\langle 12,3\right\rangle$) and then computing the denominator roots of $\fwl{H(z)}$, the following set of poles is obtained: $J=\{1.3090, 0.5000, 0.1910\}$, where set $J$ has one root with modulus greater than one, which confirms that using $\left\langle 12,3\right\rangle$, as fixed-point format, the verified system becomes unstable.

\begin{figure}[htb]
\center
\subfigure[ref1][Successful verification using the format $\left\langle 2,13\right\rangle$.]{\includegraphics[width=7cm]{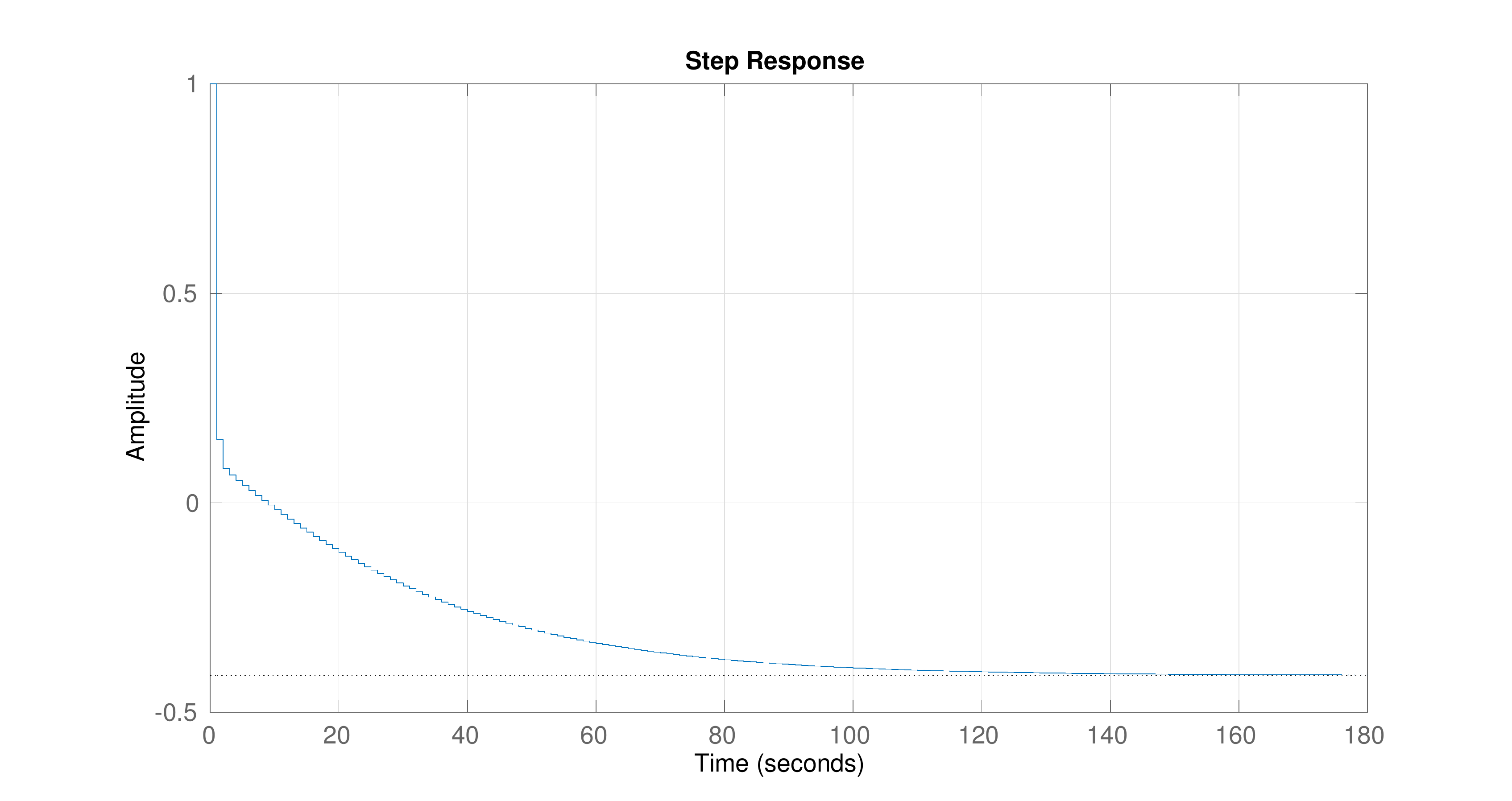}}
\qquad
\subfigure[ref2][Failed verification using the format $\left\langle 12,3\right\rangle$.]{\includegraphics[width=7cm]{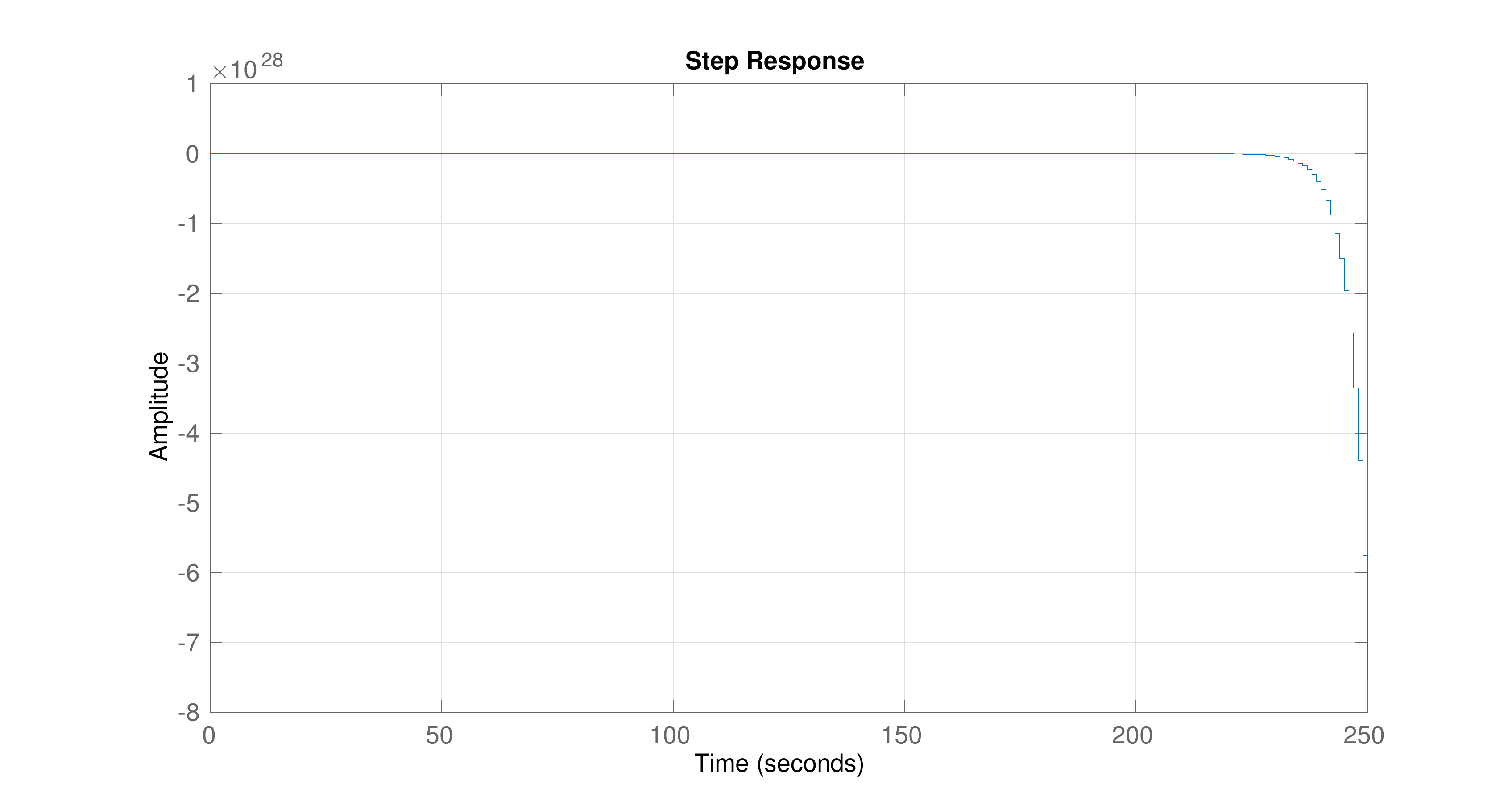}}
\caption{Step Response for Eq.~\eqref{equation_controller}.}
\label{step_response}
\end{figure}

The stability for the digital systems described above could be indeed observed through the step response for both cases, as shown in Fig.~\ref{step_response}. In subfigure \ref{step_response}(a), the step response shows that the digital system is stable, while in \ref{step_response}(b) it is unstable.

\subsubsection{GUI Application Version} 

A graphical user interface application was developed, in order to favor digital-system verification in MATLAB, besides improving usability and, consequently, attracting more digital-system engineers. Users can provide all required parameters for digital-system verification: digital-system specification, target implementation, and properties to be checked. 

\section{Conclusion}

\tool is able to verify dynamic digital-systems (controllers or filters) designed in MATLAB, through transfer-function and state-space representations in open- or closed-loop format. Regarding transfer-function representations, users are able to verify stability, minimum-phase, limit-cycle, overflow, and quantization error properties, while in state-space format, stability, quantization error, observability, and controllability properties can be verified.

We have shown that a digital controller using different numerical representations can present distinct verification results. In particular, we demonstrated that a specific representation has the potential to cause instability and then compromise the entire system's operation. \tool can verify digital systems with different implementation aspects. Given the current knowledge in formal verification, there is no other MATLAB toolbox for verifying specific properties of digital systems, while taking into account implementation aspects. 

As future work, \tool could perform verification for robust stability, by considering uncertainty in the plant and controller of closed-loop systems, and it could also be integrated into DSValidator~\cite{dsvalidator}.


\begin{thebibliography}{12}

\bibitem{diniz}
Diniz P., da~Silva E., Netto S. (2010) ``{Digital Signal Processing: System Analysis and Design}''. E-Libro, Cambridge University Press.

\bibitem{daes20161}
Bessa I. and {\it et al.} (2016)
\newblock ``{Verification of fixed-point digital controllers using direct and delta forms realizations.}''
\newblock In {\em DAES.}, 20(2):95--126.

\bibitem{peterchev}
Peterchev A and {\it et al.} (2003) Quantization resolution and limit cycling in digitally controlled {PWM} converters. In {\em IEEE Trans. Power Electronics}, 18(1): pp.301--308. 

\bibitem{peretzLC_ressonant}
Peretz M and {\it et al.} (2010) Digital Control of Resonant Converters: Resolution Effects on Limit Cycles. In {\em IEEE Trans. Power Electronics}, 25(6): pp. 1652--1661. 
  
\bibitem{bhattacharyya}
Keel L, Bhattacharyya S (1997) Robust, fragile, or optimal?. In {\em IEEE Trans. Automatic Control}, 42(8): pp. 1098--1105. 

\bibitem{spindsverifier}  Ismail H. and {\it et al.} (2015) ``{DSVerifier: A Bounded Model Checking Tool for Digital Systems}''. In {\em SPIN}, LNCS 9232, pp. 126--131. 

\bibitem{Bessa16}
Bessa I and {\it et al.} (2017) ``{Formal non-fragile stability verification of digital control systems with uncertainty}''. In {\em IEEE Trans. Computers}, 66(3): pp. 545--552.

\bibitem{monteiro2016}
Monteiro, F. R. (2016) ``{Bounded Model Checking of State-Space Digital Systems}''. In {\em FSE}, pp. 1151--1153.

\bibitem{matlab-toolbox} Matlab Toolbox (2017). In \url{https://www.mathworks.com/products/}.



\bibitem{cbmc}
Kroening, D. and Tautschnig, M. (2014) ``{CBMC -- C Bounded Model Checker},'' In {\em TACAS}, LNCS 8413, pp. 389--391.

\bibitem{esbmc}
Cordeiro and {\it et al.} (2012) ``{SMT-Based Bounded Model Checking for Embedded ANSI-C Software},'' In {\em TSE}, 38(4): pp. 957--974.

\bibitem{dsvalidator}
Chaves, L. and {\it et al.} (2016) ``{DSValidator: An Automated Reproducibility Tool for Digital Systems}'', In Technical Report published as an arXiv Document.

\end{thebibliography}
\end{document}